\documentclass[aps,pra,twocolumn,showpacs,groupedaddress]{revtex4-1}
\usepackage{amsfonts}
\usepackage{amssymb}
\usepackage{amsmath}
\usepackage{graphicx}

\begin{document}

\title{Criticality and hidden criticality in multi-species Bose-Einstein
condensates}
\author{Y.M.Liu$^{1,3}$, Y.Z.He$^2$, and C.G.Bao$^{2,}$}
\thanks{Corresponding author: stsbcg@mail.sysu.edu.cn}

\begin{abstract}
A general approach is proposed to solve the coupled Gross-Pitaevskii
equations (CGP) for K-species Bose-Einstein condensates (BEC). Analytical
solutions have been obtained under the Thomas-Fermi approximation. We aim at
finding out the common features of the K-species BEC. In particular, two
types of phase-transitions, full-state-transition and partial-state
transition, are found. In the former all species are involved in the
transition, while in the latter only a few specified species are essentially
involved. This leads to the criticality and the hidden criticality
(previously found in multi-band superconductivity). We further found that
the former originates from the singularity of the whole matrix of the CGP,
while the latter originates from the singularity of a specified sub-matrix
(which is contributed by only a few specified species). It is emphasized
that the singularity is not a by-product of the TFA, but is inherent in the
CGP.
\end{abstract}

\maketitle

\affiliation{$^1$Department of Physics, Shaoguan University, Shaoguan,
512005, P. R. China}
\affiliation{$^2$School of
Physics, Sun Yat-Sen University, Guangzhou, P. R. China}
\affiliation{$^3$State Key Laboratory of Theoretical Physics, Institute of Theoretical
Physics, Chinese Academy of Sciences, Beijing, 100190, China}

\section{Introduction}

The Bose-Einstein condensates (BEC) are artificial and controllable
many-body systems. In principle, their behavior can be designed. Therefore,
these systems are very valuable in both academic and practical senses. The
research into the 2-species BEC was initiated in 1996. Since then, an
increasing effort is paid in both the experimental aspect \cite%
{myat97,ande05,ni08,pilc09,nemi09,wack15,mgr} and theoretical aspect\cite%
{ho96,esry97,pu98,chui99,tripp00,ribo02,chui03,luo07,luo08,nott15,
scha15,inde15,kuop15,polo15,jsy,mpe,rci,jpb}. During the progress a
noticeable point is the relation between the BEC and other many-body
systems. Recently, the similarity on the criticality and the hidden
criticality between the multi-species BEC and the multi-band
superconductivity was primary found \cite{sr,sup1,sup2,sup3,sup4}. It is
believed that new and rich physical phenomena would emerge from
multi-species BEC. Some of them and those found in other many-body systems
might have the same physical background. Thus the knowledge extracted from
the multi-species BEC would be in general helpful for understanding the
behavior of other many-body systems. Moreover, multi-species BEC is
experimentally realizable. Thus, the study of this topic is meaningful and
practical. In fact, the study on 3-species BEC has already been initiated
\cite{cal,man,orl,v2}.

This paper is one along this line. it is a generalization of our two
previous paper \cite{jpa,sr} from three-species BEC to K-species, and from
adopting a spherical-symmetric trap to a more general trap (spherical- and
axial-symmetric traps are considered as special cases). The aim is to find
out the common features existing in various multi-species BEC. The emphasis
is placed on the qualitative aspect and on the critical phenomena. For this
purpose, we have to solve the coupled Gross-Pitaevskii equations (CGP) to
obtain the solution for the ground states. It is assumed that the particle
numbers are huge so that the Thomas-Fermi approximation (TFA) is reasonable
and therefore can be adopted. Under the TFA, an approach is proposed so that
the solutions are obtained with an analytical form. The analytical formalism
facilitate greatly the related analysis so that the inherent physics can be
better understood.

\section{Coupled Gross-pitaevskii equations and the formal solutions}

There are $N_{I}$ I-th-atoms, each has mass $m_{I}$ and they are interacting
via $V_{I}=c_{I,I}\underset{1\leq i<i^{\prime }\leq N_{I}}{\Sigma }\delta (%
\mathbf{r}_{i}-\mathbf{r}_{i^{\prime }})$, where $I$\ is from 1 to $K(\geq
2) $. All the particle numbers $N_{I}$ are assumed to be huge (say, $\geq $%
10000). The interspecies interactions are $V_{IJ}=c_{I,J}\underset{1\leq
i\leq N_{I}}{\Sigma }\cdot \underset{1\leq j\leq N_{J}}{\Sigma }\delta (%
\mathbf{r}_{i}-\mathbf{r}_{j})$. These atoms are confined by 3-D harmonic
traps $\frac{1}{2}m_{I}(\omega _{Ix}^{2}x^{2}+\omega _{Iy}^{2}y^{2}+\omega
_{Iz}^{2}z^{2})$. We introduce a mass $m$ and a frequency $\omega $. Then, $%
\hbar \omega $ and $\lambda \equiv \sqrt{\hbar /(m\omega )}$ are used as
units for energy and length. The spin degrees of freedom are considered as
being frozen. The total Hamiltonian is
\begin{eqnarray}
H &=&\underset{1\leq I\leq K}{\Sigma }H_{I}+\underset{1\leq I<J\leq K}{%
\Sigma }V_{IJ}  \notag \\
H_{I} &=&\sum_{i=1}^{N_{I}}[-\frac{m}{2m_{I}}\nabla _{i}^{2}+\frac{1}{2}%
(\gamma _{Ix}x_{i}^{2}+\gamma _{Iy}y_{i}^{2}+\gamma _{Iz}z_{i}^{2})]
\label{eq1} \\
&&+V_{I}
\end{eqnarray}%
where $\gamma _{Il}=(m_{I}/m)(\omega _{Il}/\omega )^{2}$, $l=x$, $y$, or $z$
(the same in the follows).

We are interested in the g.s. where no spatial excitations are involved, and
each kind of atoms are fully condensed into a state which is most
advantageous for binding. Accordingly, the total many-body wave function of
the g.s. can be written as
\begin{equation}
\Psi =\prod_{i_{1}=1}^{N_{1}}\phi _{1}(\mathbf{r}_{i_{1}})\cdot \cdot \cdot
\cdot \prod_{i_{K}=1}^{N_{K}}\phi _{K}(\mathbf{r}_{i_{K}})  \label{eq2}
\end{equation}%
where $\phi _{I}(\mathbf{r})$ is the normalized single particle wave
function. The associated CGP is a set of K coupled equations. The I-th of
them is
\begin{equation}
\lbrack -\frac{m}{2m_{I}}\nabla ^{2}+\frac{1}{2}\Sigma _{l}\gamma _{Il}l^{2}+%
\underset{1\leq J\leq K}{\Sigma }\alpha _{IJ}\phi _{J}^{2}-\varepsilon
_{I}]\phi _{I}=0  \label{eq3}
\end{equation}%
where the summation of $l$\ covers $x$, $y$, and $z$. $\alpha _{IJ}\equiv
N_{J}c_{I,J}$ and is called the weighted strength (W-strength). $\varepsilon
_{I}$ is the chemical potential. It is emphasized that the normalization ${%
\int \phi _{l}^{2}d\mathbf{r}=1}$ is required.

Since the total kinetic energy is proportional to the total particle numbers
$N$ while the total interaction energy is $\propto N^{2}$, the relative
importance of the former becomes very small when $N$\ becomes very large. In
this case, TFA (neglect the kinetic energies) can be adopted. The
applicability of this approximation has been evaluated via a numerical
approach given in \cite{yzhe,polo15}. Under the TFA and in a specific
spatial domain where all the $\phi _{I}$\ ($I$ from 1 to $K$)\ are nonzero,
the CGP can be written in a matrix form as

\begin{equation}
\mathfrak{[\alpha }_{IJ}\mathfrak{]}\left(
\begin{array}{c}
\phi _{1}^{2} \\
\cdot \\
\cdot \\
\phi _{K}^{2}%
\end{array}%
\right) =\left(
\begin{array}{c}
\varepsilon _{1}-\frac{1}{2}\Sigma _{l}\gamma _{1l}l^{2} \\
\cdot \\
\cdot \\
\varepsilon _{K}-\frac{1}{2}\Sigma _{l}\gamma _{Kl}l^{2}%
\end{array}%
\right)  \label{eq4}
\end{equation}%
where $[\alpha _{IJ}]$ denotes a K-rank matrix with elements $\alpha _{IJ}$.
When $[\alpha _{IJ}]$\ is nonsingular, its reverse $[\alpha _{IJ}]^{-1}$
exists. Then we have a formal solution for $\{\phi \}$ as
\begin{equation}
\phi _{I}^{2}=X_{I}-\Sigma _{l}Y_{I,l}l^{2}  \label{fi}
\end{equation}%
where $I$ is from 1 to $K$,
\begin{equation}
X_{I}=D_{X_{I}}/D  \label{xi}
\end{equation}%
$D$ is the determinant of $[\alpha _{IJ}]$. $D_{X_{I}}$ is a determinant
obtained by changing the $I-th$ column of $D$\ from $(\alpha _{1I},\cdot
\cdot \cdot \alpha _{KI})$ to $(\varepsilon _{1},\cdot \cdot \cdot
\varepsilon _{K})$.
\begin{equation}
Y_{I,l}=D_{Y_{Il}}/D  \label{yil}
\end{equation}%
$D_{Y_{Il}}$ is also a determinant obtained by changing the $I-th$ column of
$D$\ to $(\frac{1}{2}\gamma _{1l},\cdot \cdot \cdot \frac{1}{2}\gamma _{Kl})$%
. The set of $K$ wave functions given by eq.(\ref{fi}) is denoted as $\{\phi
\}$\ and is called a formal solution in Form K. From the reverse of eq.(\ref%
{xi}), we have a useful relation as
\begin{equation}
\varepsilon _{J}=\Sigma _{I}\alpha _{JI}X_{I}  \label{e}
\end{equation}

If in a specific spatial domain only $K^{\prime }$ $(<K)$\ wave functions\
are nonzero, then we define a subset $\sigma ^{\prime }$ of the indexes so
that all $\phi _{I^{\prime }}\geq 0$\ ($I^{\prime }\in \sigma ^{\prime }$)
and all $\phi _{J^{\prime }}=0$\ ($J^{\prime }\notin \sigma ^{\prime }$).
For the subset $\sigma ^{\prime }$ with the species number $K^{\prime }$,
the associated CGP can be written as

\begin{equation}
\mathfrak{[\alpha }_{I^{\prime }I"}\mathfrak{]}\left(
\begin{array}{c}
\phi _{1^{\prime }}^{2} \\
\cdot \\
\cdot \\
\phi _{K^{\prime }}^{2}%
\end{array}%
\right) =\left(
\begin{array}{c}
\varepsilon _{1^{\prime }}-\frac{1}{2}\Sigma _{l}\gamma _{1^{\prime }l}l^{2}
\\
\cdot \\
\cdot \\
\varepsilon _{K^{\prime }}-\frac{1}{2}\Sigma _{l}\gamma _{K^{\prime }l}l^{2}%
\end{array}%
\right)  \label{eq5}
\end{equation}%
where $[\mathfrak{\alpha }_{I^{\prime }I"}]$ denotes a $K^{\prime }$-rank
matrix with elements $\alpha _{I^{\prime }I"}$ (both $I^{\prime }$ and $%
I"\in \sigma ^{\prime }$ ), and $1^{\prime }$\ to $K^{\prime }$ denote the
indexes in $\sigma ^{\prime }$. When $[\mathfrak{\alpha }_{I^{\prime }I"}]$\
is nonsingular, we have a formal solution as
\begin{eqnarray}
(\phi _{I^{\prime }}^{\sigma ^{\prime }})^{2}=X_{I^{\prime }}^{\sigma
^{\prime }}-\Sigma _{l}Y_{I^{\prime }l}^{\sigma ^{\prime }}l^{2}(\phi
_{J^{\prime }}^{\sigma ^{\prime }})^{2} &=&0  \label{fip} \\
(\phi _{J^{\prime }}^{\sigma ^{\prime }})^{2} &=&0
\end{eqnarray}%
where%
\begin{equation}
X_{I^{\prime }}^{\sigma ^{\prime }}=D_{\sigma ^{\prime },X_{I^{\prime
}}}/D_{\sigma ^{\prime }}  \label{xis}
\end{equation}%
$D_{\sigma ^{\prime }}$ is the determinant of $[\mathfrak{\alpha }%
_{I^{\prime }I"}]$. $D_{\sigma ^{\prime },X_{I^{\prime }}}$ is a determinant
obtained by changing the $I^{\prime }-th$ column of $D_{\sigma ^{\prime }}$\
from $(\alpha _{1^{\prime }I^{\prime }},\cdot \cdot \cdot \alpha _{K^{\prime
}I^{\prime }})$ to $(\varepsilon _{1^{\prime }},\cdot \cdot \cdot
\varepsilon _{K^{\prime }})$.
\begin{equation}
Y_{I^{\prime }l}^{\sigma ^{\prime }}=D_{\sigma ^{\prime },Y_{I^{\prime
}l}}/D_{\sigma ^{\prime }}  \label{yils}
\end{equation}%
$D_{\sigma ^{\prime },Y_{I^{\prime }l}}$ is also a determinant obtained by
changing the $I^{\prime }-th$ column of $D_{\sigma ^{\prime }}$\ to $(\frac{1%
}{2}\gamma _{1^{\prime }l},\cdot \cdot \cdot \frac{1}{2}\gamma _{K^{\prime
}l})$. The set of $K$ wave functions given by eq.(\ref{fip}) are denoted as $%
\{\phi ^{\sigma ^{\prime }}\}$, and is called a formal solution in Form $%
\sigma ^{\prime }$.

From eq.(\ref{xis}) we obtain a useful formula
\begin{equation}
\varepsilon _{J^{\prime }}=\Sigma _{I^{\prime }}\alpha _{J^{\prime
}I^{\prime }}X_{I^{\prime }}^{\sigma ^{\prime }}
\end{equation}

where $I^{\prime }\in \sigma ^{\prime }$ and $J^{\prime }\in \sigma ^{\prime
}$.

Note that all the coefficients $Y_{I^{\prime }l}^{\sigma ^{\prime }}$
involved in the formal solutions are completely determined by the parameters
$\gamma _{I^{\prime }l}$ and $\alpha _{I^{\prime }J^{\prime }}$. When $%
Y_{I^{\prime }l}^{\sigma ^{\prime }}$ is positive (negative), $\phi
_{I^{\prime }}^{\sigma ^{\prime }}$ must descend (rise up) when $l$\
increases. Thus their signs are crucial to the behavior of the wave
functions.

If in a domain only one wave function, say, $\phi _{I^{\prime \prime }}$, is
nonzero, then, from eq.(\ref{eq3}) (with the kinetic term removed), it is
straight forward to obtain
\begin{equation}
\phi _{I^{^{\prime \prime }}}^{2}=\frac{1}{\alpha _{I^{\prime \prime
}I^{\prime \prime }}}(\varepsilon _{I^{\prime \prime }}-\frac{1}{2}\Sigma
_{l}\gamma _{I^{\prime \prime }l}l^{2})  \label{f1pp}
\end{equation}%
while all the other wave functions are zero. This solution is called a
formal solution in Form $1(I^{\prime \prime })$.

In what follows we shall see that these formal solutions, each hold in a
specific domain, will act as building blocks and will link up to form an
entire solution of the CGP.

\section{Two features of the formal solutions}

There are two features important to the linking of the formal solutions

\textit{Feature I: Continuity in the linking}

Let $P$\ denotes a spatial point (or a group of points, say, a curve). Let
us introduce the notation $\{\phi ^{\sigma ^{\prime }}\}_{P}$\ which implies
that the set of $K$ wave functions are all evaluated at $P$. Let $\{\sigma
^{\prime }+J^{\prime }\}$ denotes an enlarged set including all the indexes
in $\sigma ^{\prime }$\ together with an extra index $J^{\prime }\notin
\sigma ^{\prime }$. Then, for the point(s) $P$ where $\phi _{J^{\prime
}}^{\{\sigma ^{\prime }+J^{\prime }\}}=0$, we can prove that $\{\phi
^{\{\sigma ^{\prime }+J^{\prime }\}}\}_{P}=\{\phi ^{\sigma ^{\prime }}\}_{P}$
(namely, for any index $I$ (from 1 to $K$), $\phi _{I}^{\{\sigma ^{\prime
}+J^{\prime }\}}|_{P}=\phi _{I}^{\sigma ^{\prime }}|_{P}$ holds ,i.e., the
two sets of wave functions are one-by-one equal at $P$).\ This is because
the two sets of equations for $\{\phi ^{\{\sigma ^{\prime }+J^{\prime }\}}\}$
and $\{\phi ^{\sigma ^{\prime }}\}$, respectively, will become the same at $%
P $ (refer to eq.(\ref{eq5})). Thus, when the Form $\sigma ^{\prime }$\ is
transformed to $\{\sigma ^{\prime }+J^{\prime }\}$ at $P$, all the wave
functions remain to be continuous.

\textit{Feature II: When the coordinate moves along a direction, at least
one of the nonzero wave functions are descending}

This common feature holds when all the interactions are repulsive. The proof
is as follows.

From eq.(\ref{eq5}) we know that the Form $\sigma ^{\prime }$ could exist
only if the matrix $[\alpha _{I^{\prime }J^{\prime }}]$ is nonsingular.
Therefore, when each of its column is considered as a vector, the column
vectors are linearly independent. Thus, the vector $(\gamma _{1^{\prime
}l},\cdot \cdot \cdot \gamma _{K^{\prime }l})$ can be in general expanded as

\begin{equation}
\left(
\begin{array}{c}
\gamma _{1^{\prime }l} \\
\cdot \\
\cdot \\
\gamma _{K^{\prime }l}%
\end{array}%
\right) =\underset{J^{\prime }}{\Sigma }\eta _{J^{\prime }}^{(l)}\left(
\begin{array}{c}
\alpha _{1^{\prime }J^{\prime }} \\
\cdot \\
\cdot \\
\alpha _{K^{\prime }J^{\prime }}%
\end{array}%
\right)  \label{eq4p}
\end{equation}

In Insert this equation into $D_{\sigma ^{\prime },Y_{I^{\prime }l}}$, then
we have $Y_{I^{\prime }l}=\eta _{I^{\prime }}^{(l)}/2$. When all the
interactions are repulsive, all the matrix elements $\{\alpha _{I^{\prime
}J^{\prime }}\}$\ are positive. Furthermore, all the $\gamma _{I^{\prime }l}$
are also positive due to their definitions. Thus, when all the $\eta
_{I^{\prime }}^{(l)}$\ were negative, eq.(\ref{eq4p}) would lead to a
contradiction. Thus, for any $\sigma ^{\prime }$ and $l$, $Y_{I^{\prime
}l}^{\sigma ^{\prime }}$ ($I^{\prime }\in \sigma ^{\prime }$) can not all be
negative. It implies that at least one of them is positive. Accordingly, at
least one of the wave functions are descending when $l$\ increases, thus
this feature is proved.

The \textit{Feature II} has profound influence on the behavior of the
solutions of the CGP. For any formal solution at least one of the nonzero
wave functions will descend along a given direction, and therefore will
eventually become zero. Thus the form-transformation is inevitable as
explained in the next section. Furthermore, no wave functions can emerge
from an empty domain. This is because, if some of them (say, the $K^{\prime
} $ wave functions $\phi _{I^{\prime }}^{\sigma ^{\prime }}$ ($I^{\prime }$
from 1 to $K^{\prime }$) emerge together, all the related $Y_{I^{\prime
}l}^{\sigma ^{\prime }}$ must be negative to ensure the uprising. This
violates the above feature. If one wave function emerged singly, it would
violate the form given in eq.(\ref{f1pp}) (this form prohibits also the
uprising).

\section{The domain that supports a specific formal solution and its boundary%
}

Note that, when all the parameters ($\gamma _{Il}$ and $\alpha _{IJ}$) are
given and all the $\varepsilon _{I}$\ have been presumed, all the quantities
involved in the formal solution are known. Let the spatial domain that
supports the Form $\sigma ^{\prime }$ be denoted as $\Theta _{\sigma
^{\prime }}$. In $\Theta _{\sigma ^{\prime }}$ (it is called the inherent
domain of $\sigma ^{\prime }$) all the $K^{\prime }$\ inequalities $%
X_{I^{\prime }}^{\sigma ^{\prime }}-\Sigma _{l}Y_{I^{\prime }l}^{\sigma
^{\prime }}l^{2}\geq 0$ are fulfilled to ensure that all the $\phi
_{I^{\prime }}^{\sigma ^{\prime }}\geq 0$ ($I^{\prime }\in \sigma ^{\prime }$%
), while all the $\phi _{J^{\prime }}^{\sigma ^{\prime }}=0$ ($J^{\prime
}\notin \sigma ^{\prime }$) are given as zero. Obviously, $\Theta _{\sigma
^{\prime }}$ is bounded by a number of surfaces each is specified by an
equation $X_{I^{\prime }}^{\sigma ^{\prime }}-\Sigma _{l}Y_{I^{\prime
}l}^{\sigma ^{\prime }}l^{2}=0$. Among these $K^{\prime }$ surfaces, only
the inmost segments of surfaces are effective, and they constitute the whole
boundary for $\Theta _{\sigma ^{\prime }}$, thereby the domain of $\Theta
_{\sigma ^{\prime }}$\ is well defined. Each segment is called an inherent
segment. Note that $\phi _{I^{\prime }}^{\sigma ^{\prime }}$ becomes zero at
the segment specified by $X_{I^{\prime }}^{\sigma ^{\prime }}-\Sigma
_{l}Y_{I^{\prime }l}^{\sigma ^{\prime }}l^{2}=0$, therefore the neighboring
domain does not contain the $I^{\prime }$-species and is therefore denoted
as $\Theta _{\{\sigma ^{\prime }-I^{\prime }\}}$. Thus, the crossing over
the boundary causes a form-transformation. In this way the Form $\sigma
^{\prime }$ transforms to Form $\{\sigma ^{\prime }-I^{\prime }\}$ and they
are linked up continuously.

However, in an entire solution of the CGP, the actual domain with the Form $%
\sigma ^{\prime }$ (denoted as $\Omega _{\sigma ^{\prime }}$) is not
necessary to fill up $\Theta _{\sigma ^{\prime }}$. For any $J^{\prime
}\notin \sigma ^{\prime }$, when $\Theta _{\{\sigma ^{\prime }+J^{\prime
}\}} $\ contains an inherent segment\ specified by $\phi _{J^{\prime
}}^{\{\sigma ^{\prime }+J^{\prime }\}}=X_{J^{\prime }}^{\{\sigma ^{\prime
}+J^{\prime }\}}-\Sigma _{l}Y_{J^{\prime }l}^{\{\sigma ^{\prime }+J^{\prime
}\}}l^{2}=0$, then, for any point $P$ at the segment, $\{\phi ^{\{\sigma
^{\prime }+J^{\prime }\}}\}_{P}=\{\phi ^{\sigma ^{\prime }}\}_{P}$ holds
(refer to \textit{Feature I}). It implies that this segment of $\Theta
_{\{\sigma ^{\prime }+J^{\prime }\}}$ is embedded in $\Theta _{\sigma
^{\prime }}$. This embedded segment can be optionally adopted (or not
adopted, see below) as a part of the boundary for $\Omega _{\sigma ^{\prime
}}$. If it is adopted, $\Omega _{\sigma ^{\prime }}$ will be smaller than $%
\Theta _{\sigma ^{\prime }}$ and the neighboring domain by the embedded
segment will be $\Omega _{\{\sigma ^{\prime }+J^{\prime }\}}$. In general,
in addition to the inherent segments, a few embedded segments can be
optionally adopted as a part of the boundary for $\Omega _{\sigma ^{\prime
}} $. In other words, the actual domain for the Form $\sigma ^{\prime }$\
can be partially designed.

\section{Linking of the formal solutions}

When two formal solutions are linked up via a common boundary, a necessary
requirement is the continuity. This requirement is ensured due to \textit{%
Feature I}. When two segments (each is a part of a surface) intersect at a
curve, attention should be paid to the neighborhood of the curve.

\begin{figure}[tbp]
\centering \resizebox{0.95\columnwidth}{!}{\includegraphics{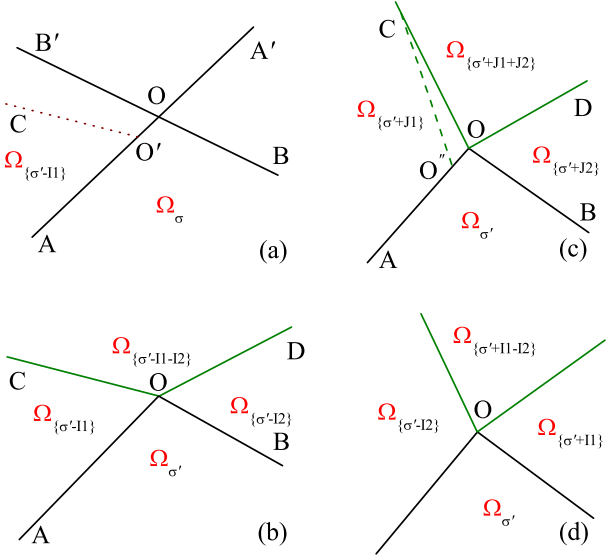} }
\caption{A draft to demonstrate the possible neighboring domains of $\Omega
_{\protect\sigma ^{\prime }}$ plotted in the r-z plane.}
\label{Fig.1}
\end{figure}

(1) When two inherent segments of $\Omega _{\sigma ^{\prime }}$\ (specified
by $\phi _{I_{1}}^{\sigma ^{\prime }}=0$ and $\phi _{I_{2}}^{\sigma ^{\prime
}}=0$) intersect at a curve $C_{I_{1}I_{2}}$, we introduce a draft shown in
Fig.1a, which is plotted in the X-Z plane while $y$\ is fixed. Where, the
two inherent segments are marked as $\overset{\_\_}{AOA^{\prime }}$\ and $%
\overset{\_\_}{BOB^{\prime }}$, the curve $C_{I_{1}I_{2}}$ is marked as a
point $O$, and the domain $\Omega _{\sigma ^{\prime }}$ is bound by $\overset%
{\_\_}{AOB}$. The domain by the other side of $\overset{\_\_}{AO}$\ is $%
\Omega _{\{\sigma ^{\prime }-I_{1}\}}$. We assume that, for an index $%
I_{x}\in \{\sigma ^{\prime }-I_{1}\}$, the inherent segment of\ $\Omega
_{\{\sigma ^{\prime }-I_{1}\}}$ specified by $\phi _{I_{x}}^{\{\sigma
^{\prime }-I_{1}\}}=0$\ intersects $\overset{\_\_}{AO}$\ at $O^{\prime }$.
Let this segment be marked by $\overset{\_\_}{O^{\prime }C}$ as shown in 1a,
thus we have
\begin{equation}
\phi _{I_{x}}^{\{\sigma ^{\prime }-I_{1}\}}|_{\overset{\_\_}{O^{\prime }C}}=0
\label{equ1}
\end{equation}

On the other hand, recall that $\overset{\_\_}{AO}$ is an inherent segment
of $\Omega _{\sigma ^{\prime }}$ at which $\phi _{I_{1}}^{\sigma ^{\prime
}}=0$ but $\phi _{I_{x}}^{\sigma ^{\prime }}\neq 0$. Thus we have
\begin{equation}
\phi _{I_{x}}^{\sigma ^{\prime }}|_{\overset{\_\_}{AO}}\neq 0  \label{equ2}
\end{equation}

Since $O^{\prime }$\ is a point at $\overset{\_\_}{O^{\prime }C}$ and also
at $\overset{\_\_}{AO}$, from eqs.(\ref{equ1},\ref{equ2}) we have

$\phi _{I_{x}}^{\{\sigma ^{\prime }-I_{1}\}}|_{O^{\prime }}\neq \phi
_{I_{x}}^{\sigma ^{\prime }}|_{O^{\prime }}$. This is in contradiction with
the \textit{Feature I} unless $I_{x}=I_{2}$ and $O^{\prime }=O$ (in this
case the inequality eq.(\ref{equ2}) becomes an equality at $O$). In fact,
the inherent segment (specified by $\phi _{I_{2}}^{\{\sigma ^{\prime
}-I_{1}\}}=0$) must intersect $\overset{\_\_}{AO}$ at $O$, because $O$\ is
the only point at $\overset{\_\_}{AO}$ where $\phi _{I_{2}}^{\{\sigma
^{\prime }-I_{1}\}}$ could be zero. Thus, we have proved that the part of
boundary of $\Omega _{\{\sigma ^{\prime }-I_{1}\}}$ will be connected as $%
\overset{\_\_}{AOC}$ as shown in Fig.1b. Obviously, the domain by the other
side of $\overset{\_\_}{OC}$\ is $\Omega _{\{\sigma ^{\prime
}-I_{1}-I_{2}\}} $.

Incidentally, if one can tune the parameters so that the segment $\phi
_{J}^{\{\sigma ^{\prime }-I_{1}+J\}}=0$ ($J\notin \sigma ^{\prime }$)\ can
intersect $\overset{\_\_}{AO}$ exactly at $O$, this segment may be
optionally introduced as an embedded segment to replace the inherent segment
$\overset{\_\_}{OC}$. But this phenomenon is very improbable to occur (i.e.,
it occurs only if the parameters are given at a particular point in the
parameter-space).

With a similar argument we can deduce that the other neighbor of $\Omega
_{\sigma ^{\prime }}$ is $\Omega _{\{\sigma ^{\prime }-I_{2}\}}$, and the
inherent segment of $\Omega _{\{\sigma ^{\prime }-I_{2}\}}$ that connects $O$%
\ is the one specified by $\phi _{I_{1}}^{\{\sigma ^{\prime }-I_{2}\}}=0$
(marked by $\overset{\_\_}{OD}$ as shown in 1b). The domain bounded by $%
\overset{\_\_}{COD}$\ should be $\Omega _{\{\sigma ^{\prime }-I_{1}-I_{2}\}}$%
. Thus, the neighborhood of $C_{I_{1}I_{2}}$ is characterized by having four
surfaces converging at $C_{I_{1}I_{2}}$. During the extension by crossing
over $\overset{\_\_}{OA}$ and $\overset{\_\_}{OC}$ (or $\overset{\_\_}{OB}$
and $\overset{\_\_}{OD}$),\ the species number $K^{\prime }$ decreases by
two.

(2) When an embedded segment of $\Omega _{\sigma ^{\prime }}$\ (specified by
$\phi _{J_{1}}^{\{\sigma ^{\prime }+J_{1}\}}=0$ and marked by $\overset{\_\_}%
{AO}$ in Fig.1c) and another embedded segment (by $\phi _{J_{2}}^{\{\sigma
^{\prime }+J_{2}\}}=0$ and by $\overset{\_\_}{BO}$) intersect at a curve $%
C_{J_{1}J_{2}}$ (marked by $O$), the neighboring domains will be $\Omega
_{\{\sigma ^{\prime }+J_{1}\}}$ and $\Omega _{\{\sigma ^{\prime }+J_{2}\}}$,
respectively. Let us first consider the\ boundary of $\Omega _{\{\sigma
^{\prime }+J_{1}\}}$. We assume that\ an inherent segment specified by $\phi
_{I}^{\{\sigma ^{\prime }+J_{1}\}}=0$ ($I\in \sigma ^{\prime }$) intersects $%
\overset{\_\_}{AO}$ at $O^{\prime }$. Note that $\phi _{I}^{\{\sigma
^{\prime }+J_{1}\}}|_{\overset{\_\_}{AO}}=\phi _{I}^{\sigma ^{\prime }}|_{%
\overset{\_\_}{AO}}$ (due to \textit{Feature I}) while the latter $\phi
_{I}^{\sigma ^{\prime }}|_{\overset{\_\_}{AO}}\neq 0$ (because $\overset{\_\_%
}{AO}$\ is embedded in the interior of $\Theta _{\sigma ^{\prime }}$).
Therefore, $O^{\prime }$ can not lie at $\overset{\_\_}{AO}$. It implies
that any inherent segments of $\Omega _{\{\sigma ^{\prime }+J_{1}\}}$ can
not touch $\overset{\_\_}{AO}$, and therefore the part of boundary extending
from $\overset{\_\_}{AO}$ can only be an embedded segment.

We assume that an embedded segment $\phi _{J_{2}}^{\{\sigma ^{\prime
}+J_{1}+J_{2}\}}=0$ intersects $\overset{\_\_}{AO}$ at $O"$ as shown in 1c.
Then we have
\begin{equation}
\phi _{J_{2}}^{\{\sigma ^{\prime }+J_{1}+J_{2}\}}|_{O"}=0.  \label{equ3}
\end{equation}

Recall that the other segment of $\Omega _{\sigma ^{\prime }}$\ is marked by
$\overset{\_\_}{BO}$ at which $\phi _{J_{2}}^{\{\sigma ^{\prime }+J_{2}\}}=0$%
. Since $O"$\ is lower than $O$\ as shown in 1c, we have
\begin{equation}
\phi _{J_{2}}^{\{\sigma ^{\prime }+J_{2}\}}|_{O"}<0  \label{equ4}
\end{equation}

However, due to the continuity of the wave function of the $J_{1}-$species
at $\overset{\_\_}{O"C}$, we have

$\phi _{J_{1}}^{\{\sigma ^{\prime }+J_{1}+J_{2}\}}|_{\overset{\_\_}{O"C}%
}=\phi _{J_{1}}^{\{\sigma ^{\prime }+J_{1}\}}|_{\overset{\_\_}{O"C}}$. When
the coordinates tend to $O"$, the latter tends to zero. Thus we have $\phi
_{J_{1}}^{\{\sigma ^{\prime }+J_{1}+J_{2}\}}|_{O"}=0$. Since the wave
function of the $J_{1}-$species is zero at $O"$, the two sets $\{\phi
^{\{\sigma ^{\prime }+J_{1}+J_{2}\}}\}$ and $\{\phi ^{\{\sigma ^{\prime
}+J_{2}\}}\}$\ should be one-by-one equal at $O"$ (due to \textit{Feature I}%
). In particular, we have
\begin{equation}
\phi _{J_{2}}^{\{\sigma ^{\prime }+J_{1}+J_{2}\}}|_{O"}=\phi
_{J_{2}}^{\{\sigma ^{\prime }+J_{2}\}}|_{O"}.  \label{equ5}
\end{equation}

The above three equations eqs.(\ref{equ3},\ref{equ4},\ref{equ5}) lead to a
contradiction, unless $O"$ and $O$\ overlap. Thus we have proved that the
extension of $\overset{\_\_}{AO}$ is the embedded segment $\overset{\_\_}{OC}
$ (specified by $\phi _{J_{2}}^{\{\sigma ^{\prime }+J_{1}+J_{2}\}}=0$), and
the part of boundary of $\Omega _{\{\sigma ^{\prime }+J_{1}\}}$ is $\overset{%
\_\_}{AOC}$ as shown in 1c. During the crossing over $\overset{\_\_}{AO}$
and $\overset{\_\_}{OC}$, $\Omega _{\sigma ^{\prime }}\rightarrow \Omega
_{\{\sigma ^{\prime }+J_{1}\}}\rightarrow \Omega _{\{\sigma ^{\prime
}+J_{1}+J_{2}\}}$. As before, one can tune the parameters so that the
segment $\phi _{J_{3}}^{\{\sigma ^{\prime }+J_{1}+J_{3}\}}=0$\ ($J_{3}\neq
J_{2}$) intersects $\overset{\_\_}{AO}$ also at $O$. This phenomenon is
omitted due to the very small probability (i.e., the parameters are given at
a particular set of values).

With a similar argument, for $\Omega _{\{\sigma ^{\prime }+J_{2}\}}$, the
extension of $\overset{\_\_}{BO}$\ is also an\ embedded segment specified by
$\phi _{J_{1}}^{\{\sigma ^{\prime }+J_{1}+J_{2}\}}=0$ (marked by $\overset{%
\_\_}{OD}$). The domain by the other side of $\overset{\_\_}{OD}$\ is also $%
\Omega _{\{\sigma ^{\prime }+J_{1}+J_{2}\}}$. The domains in the
neighborhood of the intersection $C_{J_{1}J_{2}}$\ is shown in Fig.1c. Note
that, by defining $\sigma "=\sigma ^{\prime }+J_{1}+J_{2}$, 1c becomes 1b
with $\sigma ^{\prime }$ being replaced by $\sigma "$. Similarly, by
defining $\sigma ^{\prime \prime \prime }=\sigma ^{\prime }-I_{1}$, 1b can
be re-plotted as 1d where an inherent segment and an embedded segment of $%
\Omega _{\sigma ^{\prime }}$\ intersect.

Fig.1 demonstrates qualitatively how the formal solutions will link up
together so that the entire solution can extends outward from a specific
domain $\Omega _{\sigma ^{\prime }}$. Recall that there are options in the
extension to be decided in advance. When the boundary of $\Omega _{\sigma
^{\prime }}$\ contains only a single inherent segment $\phi _{I_{o}}^{\sigma
^{\prime }}=0$, the outer domain is just $\Omega _{\{\sigma ^{\prime
}-I_{o}\}}$, and $\Omega _{\sigma ^{\prime }}$\ as a whole is embraced by $%
\Omega _{\{\sigma ^{\prime }-I_{o}\}}$.

\section{An approach to obtain an entire solution}

Based on the linking of the formal solutions, we propose the following
approach to obtain the entire solutions for the CGP. Firstly we prescribe
the form of the inmost domain where the origin is included. Due to the
symmetry of the trap, we can consider only the first octant (i.e., $l\geq 0$%
). Secondly, when all the parameters $\gamma _{Il}$ and $\alpha _{IJ}$ are
given, we presume all the values of $\varepsilon _{I}$. Then,\ all the
coefficients involved in the formal solutions can be known. Thirdly, it is
reminded that, during the extension from the inmost domain outward, we have
the options to adopt embedded segments. It implies that we have the options
to choose the neighbors with the choices shown in Fig.1. Where, when two
domains are connected via a surface, the species numbers of the two differ
by one. Whereas when two domains are connected only via a curve (say, $%
\Omega _{\sigma ^{\prime }}$ and $\Omega _{\{\sigma ^{\prime
}-I_{1}-I_{2}\}} $\ in Fig.1b), the species numbers differ by two.
Nevertheless, we have to make all the options in advance. In this way we
have prescribed (designed) a specific way to link up the formal solutions
continuously to form a candidate of the entire solution. Of course, every
kind of species must be included in the design. Note that, for an inherent
segment of a domain with Form 1, the outer side of the segment is empty.
According to \textit{Feature II} no wave functions can emerge from an empty
domain. Thus this inherent segment is the boundary of the whole condensate
and the extension ends.

With a well defined design it turns out that the inputs (the parameters and
the set $\{\varepsilon _{I}\}$) are seriously limited. For an example, if
the inmost domain has been prescribed to have the Form $\sigma ^{\prime }$,\
then all the $K^{\prime }$\ coefficients $X_{I^{\prime }}^{\sigma ^{\prime
}} $\ are required to be $>0$. Accordingly, constraints are imposed via the $%
K^{\prime }$\ inequalities. In general, for each reasonable design, the
inputs are constrained by a number of inequalities so that the parameters
together with $\{\varepsilon _{I}\}$ are limited within a non-null specific
scope (examples are given below). The last step is to introduce the $K$\
equations of normalization. For a given set of parameters inside the scope,
these equations are sufficient to determine the $K$\ presumed values of $%
\{\varepsilon _{I}\}$. If the $\{\varepsilon _{I}\}$\ so determined turn out
to lie also inside the scope, then a realistic entire solution is thereby
obtained (i.e., all the normalized wave functions together with $%
\{\varepsilon _{I}\}$ are known, and all the constraints are fulfilled).
Otherwise or the scope itself is null, the design is unreasonable and should
be revised. Examples with $K=4$\ is given in the next section.

\section{Examples with K=4}

To demonstrate the realization of the above approach, two examples with $K=4$%
\ are given below. For convenience, we assume that the trap is
axial-symmetric with respect to the Z-axis, and we introduce $%
r^{2}=x^{2}+y^{2}$. As mentioned, the values of $\{\gamma _{Il}\}$, $%
\{\alpha _{IJ}\}$, and $\{\varepsilon _{I}\}$\ have been presumed so that
all formal solutions are well defined.

(1) Example 1.

Let the four kinds of atoms are denoted by I, II, III, and IV. Let $\sigma
_{4}$\ denote the set containing I to IV, $\sigma _{3}$\ denote the set
containing I to III, $\sigma _{2}$\ denote the set containing I and II,
while $\sigma _{1}$\ denote the set containing I only. The design for a
miscible state is shown in Fig.2 where the domains for the formal solutions
are plotted in the first quadrant of the $r-z$ plane.

For this miscible state analytical solution can be obtained (refer to \cite%
{jpa} for $K=3$). However, analytical expressions for $K=4$\ are very
complicated and will not be given here. Instead, we will list all the
inequalities for constraining the parameters, and we will present the
numerical result with respect to a given set of parameters.

According to the design, the inmost domain (denoted as $\Omega _{\sigma
_{4}} $) contain four kinds of atoms. Therefore it is required

(i) $X_{J}^{\sigma _{4}}>0$, $J=I$ to $IV$.

(ii) The outer boundary of $\Omega _{\sigma _{4}}$ is denoted by $S_{4}$,
which is an inherent segment specified by $\phi _{IV}^{\sigma
_{4}}|_{S_{4}}=X_{IV}^{\sigma _{4}}-Y_{IV,x}^{\sigma
_{4}}r^{2}-Y_{IV,z}^{\sigma _{4}}z^{2}=0$. At $S_{4}$, $\phi _{I}^{\sigma
_{4}}|_{S_{4}}>0$, $\phi _{II}^{\sigma _{4}}|_{S_{4}}>0$, and $\phi
_{III}^{\sigma _{4}}|_{S_{4}}>0$ are required.

(iii) The outer boundary of $\Omega _{\sigma _{3}}$ is denoted by $S_{3}$
specified by $\phi _{III}^{\sigma _{3}}|_{S_{3}}=X_{III}^{\sigma
_{3}}-Y_{III,x}^{\sigma _{3}}r^{2}-Y_{III,z}^{\sigma _{3}}z^{2}=0$. At this
surface $\phi _{I}^{\sigma _{3}}|_{S_{3}}>0$ and $\phi _{II}^{\sigma
_{3}}|_{S_{3}}>0$ are required

(iv) The outer boundary of $\Omega _{\sigma _{2}}$ is denoted by $S_{2}$\
specified by $\phi _{II}^{\sigma _{2}}|_{S_{2}}=X_{II}^{\sigma
_{2}}-Y_{II,x}^{\sigma _{2}}r^{2}-Y_{II,z}^{\sigma _{2}}z^{2}=0$. At this
surface $\phi _{I}^{\sigma _{2}}|_{S_{2}}>0$ is required

(v) The outer boundary of $\Omega _{\sigma _{1}}$ is denoted by $S_{1}$\
specified by $2\varepsilon _{1}-\gamma _{I,x}r^{2}-\gamma _{I,z}z^{2}=0$. $%
S_{1}$ is also the boundary of the whole system.

(vi) Furthermore, since $S_{4}$, $S_{3}$, and $S_{2}$ are assumed to be an
ellipsoid, it is required

$Y_{IV,x}^{\sigma _{4}}>0$ and $Y_{IV,z}^{\sigma _{4}}>0$; $%
Y_{III,x}^{\sigma _{3}}>0$ and $Y_{III,z}^{\sigma _{3}}>0$; and $%
Y_{II,x}^{\sigma _{2}}>0$ and $Y_{II,z}^{\sigma _{2}}>0$.

(vii) Finally, the four equations of normalization

$1=\int \phi _{J}^{2}d\Omega $

are required, where $\phi _{J}=\phi _{J}^{\sigma _{L}}$ if $d\Omega $ is in $%
\Omega _{\sigma _{L}}$, $J$ and $L$\ are from $I$\ to $IV$.

The parameters are given as: $N_{I}=55000$, $N_{II}=37500$, $N_{III}=25000$,
$N_{IV}=12500$. The intra-species interactions in the unit $\hbar \omega
\lambda ^{3}$ are $c_{I}=c_{II}=c_{III}=c_{IV}=10^{-3}$. All the
inter-species interactions in $\hbar \omega \lambda ^{3}$ are $c_{JJ^{\prime
}}=3\times 10^{-4}$. $\gamma _{Jx}=\gamma _{Jy}=1$ and $\gamma _{Jz}=1.5$
(where $J=I$ to $IV$). Then, the corresponding solution of the CGP has the
chemical potentials (in $\hbar \omega $) $\varepsilon _{I}=4.56$, $%
\varepsilon _{II}=4.25$, $\varepsilon _{III}=3.98$, $\varepsilon _{IV}=3.64$%
. Fig.2 demonstrates not only the design but also the exact locations of the
domains. The wave functions $\phi _{J}$\ ($J=$I to IV) are shown in Fig.3.

\begin{figure}[tbp]
\centering \resizebox{0.95\columnwidth}{!}{\includegraphics{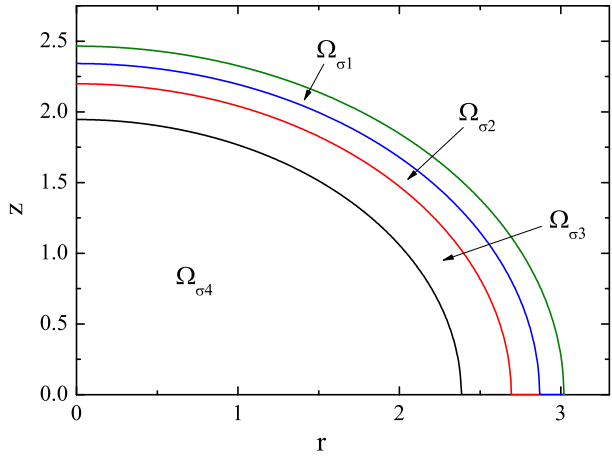} }
\caption{The distribution of the species in Example 1. The space are divided
into domains each supports a type of formal solution (plotted in the upper
r-z plane). The labels of the domains are marked. The unit for distance is $%
\protect\lambda \equiv \protect\sqrt{\hbar /(m\protect\omega )}$ (the same
in the following figures).}
\label{Fig.2}
\end{figure}

\begin{figure}[tbp]
\centering \resizebox{0.95\columnwidth}{!}{\includegraphics{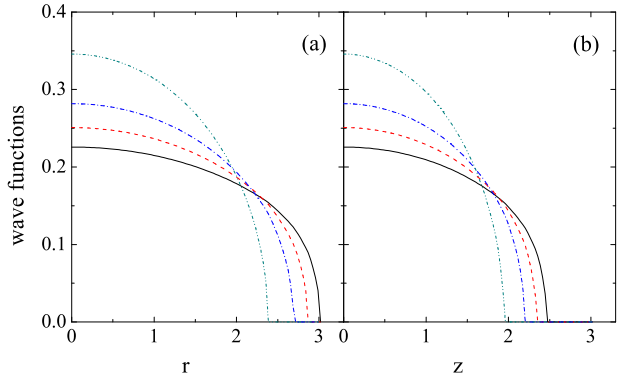} }
\caption{The wave functions in Example 1 plotted versus $r$ when $z=0$ (a),
and versus $z$ when $r=0$ (b). $\protect\phi _{I},\protect\phi _{II},\protect%
\phi _{III}$, and $\protect\phi _{IV} $ are given in solid, dashed,
dash-dot, and dash-dot-dot lines, respectively}
\label{Fig.3}
\end{figure}

(2) Example 2.

The design is shown in Fig.4. There are two intersecting inherent segments
(as in Fig.1b), denoted as $S_{4}$ (the upper boundary of $\Omega _{\sigma
_{4}}$) and $S_{4}^{\prime }$ (the right boundary of $\Omega _{\sigma _{4}}$%
), contained in the inmost domain. In addition to $\sigma _{3}$, the set $%
\sigma _{3}^{\prime }$\ is introduced which contains I, II, and IV. The
corresponding domains are $\Omega _{\sigma _{3}}$ and $\Omega _{\sigma
_{3}^{\prime }}$, their right boundaries are denoted as $S_{3}$ and $%
S_{3}^{\prime }$, respectively

The constraints imposing upon the parameters and $\{\varepsilon _{I}\}$\ are
as follows

(i) The four inequalities $X_{J}^{\sigma _{4}}>0$ hold as in Example 1.

(iia) At the segment $S_{4}$, $\phi _{I}^{\sigma _{4}}|_{S_{4}}>0$, $\phi
_{II}^{\sigma _{4}}|_{S_{4}}>0$, $\phi _{III}^{\sigma _{4}}|_{S_{4}}>0$, and
$\phi _{IV}^{\sigma _{4}}|_{S_{4}}=0$.

(iib) At the segment $S_{4}^{\prime }$, $\phi _{I}^{\sigma
_{4}}|_{S_{4}^{\prime }}>0$, $\phi _{II}^{\sigma _{4}}|_{S_{4}^{\prime }}>0$%
, $\phi _{IV}^{\sigma _{4}}|_{S_{4}^{\prime }}>0$, and $\phi _{III}^{\sigma
_{4}}|_{S_{4}^{\prime }}=0$. \ \

(iiia) At the segment $S_{3}$, $\phi _{I}^{\sigma _{3}}|_{S_{3}}>0$, $\phi
_{II}^{\sigma _{3}}|_{S_{3}}>0$, and $\phi _{III}^{\sigma _{3}}|_{S_{3}}=0$.

(iiib) At the segment $S_{3}^{\prime }$, $\phi _{I}^{\sigma _{3}^{\prime
}}|_{S_{3}^{\prime }}>0$, $\phi _{II}^{\sigma _{3}^{\prime
}}|_{S_{3}^{\prime }}>0$, and $\phi _{IV}^{\sigma _{3}^{\prime
}}|_{S_{3}^{\prime }}=0$ .

Furthermore, the points (iv), (v), (vi), and (vii) of Example 1 hold also in
Example 2. In addition, $Y_{III,x}^{\sigma _{4}}>0$, $Y_{III,z}^{\sigma
_{4}}>0$, $Y_{IV,x}^{\sigma _{3}^{\prime }}>0$ and $Y_{IV,z}^{\sigma
_{3}^{\prime }}>0$ are required.

The parameters are adopted as:

$N_{I}=62500$, $N_{II}=45000$, $N_{III}=25000$, $N_{IV}=25000$. $\gamma
_{Ix}=\gamma _{IIx}=0.8$, $\gamma _{Iz}=\gamma _{IIz}=0.7$,$\ \gamma
_{IIIx}=1.2$ and $\gamma _{IIIz}=0.8$, while $\gamma _{IVx}=0.8$ and $\gamma
_{IVz}=1.2$. For the interactions, $c_{I}=c_{II}=c_{III}=c_{IV}=10^{-3}$ and
$c_{JJ^{\prime }}=3\times 10^{-4}$ as in Example 1. Note that the III-atoms
(IV-atoms) are prescribed to be looser (tighter) bound along z-axis, thus
they can be distributed more extensive (compact) along z-axis. Then, the
corresponding solution of the CGP has $\varepsilon _{I}=4.00$, $\varepsilon
_{II}=3.78$, $\varepsilon _{III}=4.04$, $\varepsilon _{IV}=3.75$. In Fig.4,
not only the design but also the exact locations of the domains are
demonstrated. The wave functions $\phi _{J}$\ ($J=$I to IV) are shown in
Fig.5.

\begin{figure}[tbp]
\centering \resizebox{0.95\columnwidth}{!}{\includegraphics{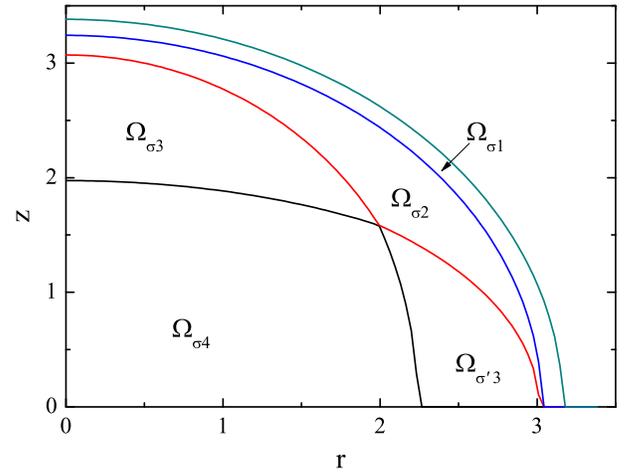} }
\caption{The same as Fig.2 but for Example 2.}
\label{Fig.4}
\end{figure}

\begin{figure}[tbp]
\centering \resizebox{0.95\columnwidth}{!}{\includegraphics{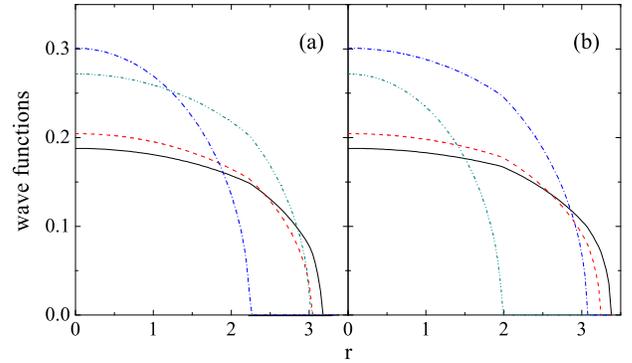} }
\caption{The same as Fig.3 but for Example 2.}
\label{Fig.5}
\end{figure}

\section{\ Common features of $K-$species condensates and the inherent
state-transition}

We have proposed an approach to solve the CGP of the $K-$species condensates
under the TFA. The traps for all the atoms have a common center. The common
features of these systems are as follows:

(i) The entire solution is composed of the formal solutions, each hold in a
specific domain and each contains specific kinds of atoms.

(ii) The domain that supports a specific formal solution can not extend
along a direction infinitively (\textit{Feature II}). Therefore,
form-transformation is inevitable and the formal solutions are thereby
linked up. During the linking of the formal solutions the wave function of
every species remains continuous (\textit{Feature I}).

(iii) No wave functions can emerge from an empty domain (due to \textit{%
Feature II}). Thus the atoms can not be distributed in disconnected regions.
Furthermore, the center of the traps can not be empty. In particular, the
inherent segment of a domain with Form 1 is the outmost boundary of the
whole system. This is because the domain next to the segment is empty.

(iv) Usually, the species numbers of two neighboring domains differ by one.
However, for $K\geq 3$, the condensate may contain a special structure in
which four domains with the Form $\sigma ^{\prime }$, $\{\sigma ^{\prime
}-I_{1}\}$, $\{\sigma ^{\prime }-I_{2}\}$, and $\{\sigma ^{\prime
}-I_{1}-I_{2}\}$, respectively, are neighboring to each other, and their
boundaries converge at a curve at which $\phi _{I_{1}}=0$ and $\phi
_{I_{2}}=0$. The convergence is shown in Fig.1 (where the set $\{\sigma
^{\prime }-I_{1}-I_{2}\}$ might be null). It arises because the four sets of
wave functions in the above four forms, respectively, will satisfy the same
matrix-equation at the curve. Thus they are one-to-one equal at the curve,
and therefore they converge. The convergence will be a popular phenomenon
(but not of necessity) in K-species BEC.

(v) There are critical phenomena inherent in the multi-species BEC. When the
strengths of interaction are tuned so that the matrix $[\alpha _{IJ}]$ tends
to be singular (i.e., its determinant $D\rightarrow 0$), the wave functions
in Form K will become extremely steep because $Y_{I,l}\rightarrow \pm \infty
$ (refer to eq.(\ref{fi},\ref{yil})). When an entire solution of the CGP\
contains a Form K, during the crossing over the singular point of $[\alpha
_{IJ}]$, $Y_{I,l}$ (for all $I$ and $l$) suddenly changes its sign and jumps
from $+\infty $ to\ $-\infty $, or vice versa. Accordingly, the whole set of
steeply down-falling wave functions suddenly become uprising, or vice versa.
This causes a great change in the composition of the state, i.e., a
state-transition. Accompanying the transition, the total energy increases
remarkably as shown in the references \cite{jpb,sr}. Since all the K species
are involved in this transition, it is called a full-state-transition.
Obviously, the coupling among all the species contributes to this critical
phenomenon. Therefore, when more and more species take part in the coupling,
the critical point (i.e., the singular point of the matrix) will shift. The
shift has been confirmed in a previous paper on 3- species BEC \cite{sr}.

Furthermore, when an entire solution\ contains a Form $\sigma ^{\prime }$
with $K^{\prime }$\ species, the singular point of the $K^{\prime }-$rank
matrix $[\alpha _{I^{\prime }J^{\prime }}]$ is also a critical point (where $%
D_{\sigma ^{\prime }}=0$). The crossing over this point will lead also to a
state-transition. However, during the transition, only the nonzero wave
functions of the $K^{\prime }-$species are essentially involved (refer to
eqs.(\ref{fip,fils})), while the other are only slightly influenced (refer
to \cite{sr}). Therefore, it is called a partial-state-transition. In this
case, the transition is strongly affected by the interactions among the $%
K^{\prime }$ species but only weakly affected by the inter-species
interaction imposed by the other $K-K^{\prime }$ species. In particular, the
critical point of the partial-state-transition of the $K$-BEC is exactly the
same as the full-state-transition of the\ $K^{\prime }$-BEC. When an entire
solution contains a Form $\sigma ^{\prime }$, the associated
partial-state-transition might occur.

Note that the equation $D_{\sigma ^{\prime }}=0$\ specifies a surface in the
parameter-space $\Sigma _{P}$. Therefore, each of the above critical point
is in fact a critical surface in $\Sigma _{P}$. Thus, $\Sigma _{P}$ is
divided into zones by a number of surfaces, one is related to the
full-state-transition while the others are related to
partial-state-transition. Each zone supports a specific phase. These zones
might be further divided to give a more detailed classification on the
composition. The division of $\Sigma _{P}$\ into zones would lead to the
phase-diagram.

Similar to the importance of the inter-species coupling in multi-species
BEC, the inter-band coupling in multi-band superconductivity are found to be
also important to the critical phenomena \cite{sup1,sup2,sup3,sup4}. Due to
the coupling the critical points are therefore shifted. Moreover, the
occurrence of the partial-state-transition can be regarded as\ a kind of
hidden criticality, namely, the criticality of a sub-system (with a specific
critical point) is hidden in the whole system and would emerge under certain
conditions. This is more or less similar to the hidden criticality found in
multi-band superconductivity \cite{sup2}.

Recall that, in the earliest study of the 2-species BEC, it has already been
pointed out that the singular point of the two-rank matrix of the CGP
induces instability \cite{ho96}. Obviously, the singularity of the matrix is
an inherent feature of the CGP irrelevant to the TFA (refer to the last
section of the ref. \cite{sr}). Therefore, the above state-transition is not
a by-product of the TFA but an inherent physical phenomenon common to all $K-
$BEC.

\begin{acknowledgments}
Supported by the National Natural Science Foundation of China under Grants
No.11372122, 11274393, 11574404, and 11275279; the Open Project Program of
State Key Laboratory of Theoretical Physics, Institute of Theoretical
Physics, Chinese Academy of Sciences, China(No.Y4KF201CJ1); and the National
Basic Research Program of China (2013CB933601).
\end{acknowledgments}

\end{document}